\documentclass{PoS}
\usepackage{cite} 
\usepackage{amsmath}

\newcommand{\be}{\begin{equation}}
\newcommand{\ee}{\end{equation}}

\title{Comments on the double copy construction for gravitational theories}

\ShortTitle{Double copy construction}

\author{\speaker{Gabriel Lopes Cardoso} \\
        Center for Mathematical Analysis, Geometry and Dynamical Systems,\\
Department of Mathematics, Instituto Superior T\'ecnico,
Universidade de Lisboa, \\Av. Rovisco Pais, 1049-001 Lisboa, Portugal \\
        E-mail: \email{gcardoso@math.tecnico.ulisboa.pt}}

\author{Gianluca Inverso
\thanks{Preprint number  QMUL-PH-18-04}\\
                Centre for Research in String Theory,\\
        School of Physics and Astronomy, 
        Queen Mary, University of London,\\
         327 Mile End Road, London, E1 4NS, United Kingdom\\
        E-mail: \email{g.inverso@qmul.ac.uk}}

\author{Silvia Nagy\\
              School of Physics and Astronomy, University of Nottingham, \\
        Nottingham NG7 2RD, United Kingdom\\
        E-mail: \email{Silvia.Nagy1@nottingham.ac.uk }}

\author{Suresh Nampuri\\
        Center for Mathematical Analysis, Geometry and Dynamical Systems,\\
Department of Mathematics, Instituto Superior T\'ecnico,
Universidade de Lisboa, \\ Av. Rovisco Pais, 1049-001 Lisboa, Portugal \\
        E-mail: \email{nampuri@gmail.com}}

\abstract{
We revisit the double copy description for
linearized gravity and point out various technical issues and subtleties,
associated with setting up the double copy description, including the problem of matching degrees of freedom on both sides of
the double copy dictionary and the related issue of the constraint between 
graviton and dilaton sources. We introduce and discuss possible resolutions of these issues.}

\FullConference{Corfu Summer Institute 2017 'School and Workshops on Elementary Particle Physics and Gravity'\\
		2-28 September 2017\\
		Corfu, Greece}

\begin{document}

\section{Introduction}

The double copy construction \cite{Adamo:2017nia,Anastasiou:2014qba,Anastasiou:2015vba,Borsten:2013bp,Anastasiou:2013hba,Anastasiou:2016csv,Anastasiou:2017nsz,Borsten:2017jpt,Bianchi:2008pu,Bern:2008qj,Bern:2010ue,Bern:2017yxu,Bern:2017ucb,Bern:2015ooa,Borsten:2015pla,Cardoso:2016ngt,Cardoso:2016amd,Chiodaroli:2014xia,Chiodaroli:2016jqw,Chiodaroli:2017ngp,Chiodaroli:2017ehv,Carrasco:2015iwa,Chiodaroli:2015rdg,Carrasco:2016ldy,Chester:2017vcz,Carrillo-Gonzalez:2017iyj,DeSmet:2017rve,Geyer:2017ela,Goldberger:2016iau,Goldberger:2017frp,Goldberger:2017ogt,Johansson:2017srf,Luna:2015paa,Luna:2016due,Luna:2016hge,Luna:2017dtq,Monteiro:2014cda,Monteiro:2015bna,White:2017mwc,
Cachazo:2013iea,Goldberger:2017vcg,Chiodaroli:2015wal,Johansson:2014zca,Naculich:2014naa,Li:2018qap} is predicated on the premise that on-shell gravitational configurations
can be described in terms of two copies of on-shell field theory configurations.  These configurations 
are viewed as describing fluctuations around 
a fixed background. In what follows, we will restrict to linearized fluctuations, in terms of the expansion 
parameter $1/M_{Pl}$, and take the fixed background to be the Minkowski spacetime $\mathbb{R}^{1,3}$.  Specifically, in position space, 
the double copy dictionary expresses a gravitational field configuration as 
a sum of convolutions $\star$  of two copies of field theoretic configurations \cite{Anastasiou:2014qba,Cardoso:2016ngt,Cardoso:2016amd}. 
For linearized gravity, the dictionary takes the schematic form,
\begin{equation}
\varphi_G = \varphi \star {\tilde \varphi} \;,
\label{dcdic}
\end{equation}
where on the left hand side $\varphi_G$ denotes a gravitational field configuration, while on the right
hand side $\varphi$ and $\tilde \varphi$ denote field theory configurations.

There are two questions that are essential to frame a discussion of the double copy formalism, when considering (\ref{dcdic}):
\begin{enumerate}
\item What is the class of functions for which the convolution is well defined, and what are 
the ensuing convolution properties?

\item What is the class of gravitational theories that lend themselves to a double copy construction- and what are the necessary characteristics for a field theory, if it has to be one of the two copies in the double copy construction of a gravitational theory?
\end{enumerate}

In this note, we will attempt to shed some light on various aspects of  these questions. In particular, we will clarify how to ensure the matching of the degrees of freedom on both sides of the relation (\ref{dcdic}) when reproducing from the double copy (the linearization of) classical solutions of the gravitational equations of motion.

\section{Convolution properties}

At linearized level, the double copy dictionary is based on the convolution $\star$ which, in Cartesian
coordinates on $\mathbb R^{1,3}$, takes the form
\begin{equation}
[ f \star g] (x) = \int d^4 y \, f(x-y) \, g(y) \;. 
\label{conv}
\end{equation}
Here, $f, g$ and $f \star g$ are functions on $\mathbb{R}^{1,3}$.
In general, the convolution is well-defined only if $f$ and $g$ fall off sufficiently rapidly at infinity in order
for the integral to exist.  For instance, if $f$ and $g$ are continuous functions with compact support, then their convolution exists and describes a continuous function with compact support.

Assuming that the convolution integral (\ref{conv}) is well defined, it satisfies the property of
commutativity, $ f \star g = g \star f$, and of associativity, $f \star ( g \star h) = ( f \star g) \star h$.
In addition, one may also want to impose the derivative rule 
\begin{equation}
\partial \left( f \star g \right) = (\partial f) \star g = f \star (\partial g) \;,
\label{derrule}
\end{equation}
which may result in additional restrictions on the functions $f$ 
and $g$. For instance, in one dimension, a class of functions for which (\ref{derrule}) holds is provided
by functions $f$ and $g$ that belong to $L^1(\mathbb{R})$, with 
$df/dx$ and $dg/dx$ also belonging to $L^1(\mathbb{R})$.

In the following, when working with the convolution (\ref{conv}), we will take the functions that enter in 
(\ref{conv}) to satisfy:

\begin{itemize}
\item 
$ f \star g = g \star f$ and $f \star ( g \star h) = ( f \star g) \star h \equiv f\star g\star h$;

\item the derivative rule (\ref{derrule}).

\end{itemize}

\noindent

As already mentioned, demanding these properties to hold will, in general,
imply that the allowed functions must fall off quickly enough at infinity.
In particular, this implies that plane wave solutions satisfying $\Box F = 0$ \emph{globally} on $\mathbb R^{1,3}$ are no longer allowed.
We now derive a few consequences of these restrictions and of the convolution properties we just introduced:
\begin{itemize}
\item we define a Green's function $G$ by
\be
\Box_x G(x-y) = \delta^{(4)} (x-y)
\ee
and introduce the notation
\be
\frac{1}{\Box} \, (\cdot) \equiv G \star (\cdot) \;\; \;.
\label{defboxi}
\ee
Then, given a function $F$, we have
\be 
F(x) = \int \delta^{(4)}(x-y) \, F(y) d^4y= \int \Box_x  G(x-y) \, F(y) \, d^4y
= \Box \left(G \star F \right) = \Box  \left(\frac{1}{\Box} F \right)\;.
\ee
\item If we assume that $F\neq0$ commutes with $G$, i.e. $G \star F = F \star G$, we also get
\be
F = \Box \left(G \star F \right) = \Box \left(F \star G \right) = (\Box F) \star G \neq0\;,
\ee
and we conclude that $F$ can not be in the kernel of $\Box$.
We arrive at the same conclusion if we use the derivative rule:
\begin{equation}
F = \Box \left(G \star F \right)  = G \star (\Box F) \neq0\;.
\end{equation}
\item Since $F$ is not in the kernel of $\Box$, we have $\Box F=j$.
Then, using (\ref{defboxi}), we can write
\be
F = \left( \frac{1}{\Box} j \right) +\chi \equiv \left( G \star j \right) + \chi \;,
\ee
where $\chi$ belongs to the kernel of $\Box$, i.e. $\Box \chi =0$. 
\item Let us now restrict to $F$ in the image of $1/\Box$, i.e. $\chi=0$ and
\begin{equation}
F = \left( \frac{1}{\Box} j \right) = \left( G \star j \right)\;,
\end{equation}
for some function $j=\Box F$. We immediately see that for this class of functions
\begin{equation}
\frac{1}{\Box}\left(\Box F\right) = \Box \left(\frac{1}{\Box} F\right) = F\;.
\label{bbbb}
\end{equation}
Notice that this is equivalent to the derivative rule for $G\star F$:
\begin{equation}
\Box(G\star F) = (\Box G)\star F = G\star (\Box F) = F\;.
\end{equation}

\item If we also require that $F=\frac{1}{\Box} j$ commutes with $G$ and that associativity holds for convolutions of $G$ and $j$, we conclude that $j$ commutes with $G$:
\begin{equation}
G\star F-F\star G = G\star(G\star j-j\star G) = 0\;.
\end{equation}
On the other hand, if $F$ and $\Box F$ commute with $G$, then $F$ is in the image of $1/\Box$:
\be
\frac{1}{\Box} \left(\Box F \right) = G \star \left(\Box F \right) = (\Box F) \star G = \Box \left(F \star G \right)
= \Box \left(G \star F \right) =  F \;.
\label{id2}
\ee

\end{itemize}

Below, we will restrict to classes of functions for which the convolution satisfies commutativity, associativity and the derivative rule, also with respect to a Green's function $G$.
As shown above, this implies that all functions belong to the image of $1/\Box$.
We will make extensive use of (\ref{bbbb}) when discussing the double copy dictionary.
As an application, consider two one-forms $A$ and $\tilde A$ and the expression
\be
\frac{1}{\Box} \left( \partial \cdot A \star \partial \cdot {\tilde A} \right) \;.
\ee
Under gauge transformations $A \rightarrow A + d \alpha$, this shifts by a term
\be
\frac{1}{\Box} \left( \Box \alpha \star \partial \cdot {\tilde A} \right) \;.
\label{box2al}
\ee
Let us assume that also $\alpha$ is in the image of $1/\Box$.
Using associativity we immediately deduce
\be
\frac{1}{\Box} \left( \Box \alpha \star \partial \cdot {\tilde A} \right) = \alpha \star \partial \cdot {\tilde A} \;.
\label{boxgau1}
\ee
Similarly under gauge transformations ${\tilde A} \rightarrow {\tilde A} + d {\tilde \alpha}$:
we use that $\tilde\alpha$ is in the image of $1/\Box$ and also that $\partial\cdot A$ commutes with $G$.
Then,
\be
\frac{1}{\Box} \left( \partial \cdot A \star \Box {\tilde \alpha} \right) = \partial \cdot A \star  {\tilde \alpha} \;.
\label{boxgau2}
\ee
The assumptions made above are necessary to guarantee that the commutativity, associativity and derivative properties of the convolution are valid also when we perform gauge transformations.

As indicated above, we will assume that the field configurations entering the double copy description
satisfy equations of motion that are sourced, to ensure that the convolution integral
(\ref{conv}) is well defined and the convolution properties listed above hold.
This excludes configurations that describe plane waves globally on $\mathbb R^{1,3}$.

\section{Double copy dictionary}

In this note we will consider a gravitational theory based on a spacetime metric $g_{\mu \nu}$,
a massless two-form field $B_{\mu \nu}$ and a massless scalar field $\phi$. This is the universal bosonic sector of any four-dimensional low energy effective theory arising from a toroidal string compactification. 
We will
construct the double copy description of this theory at the linearized level, and discuss some of the subtleties involved.

We linearize $g_{\mu \nu}$ around flat spacetime, $g_{\mu \nu} = \eta_{\mu \nu} + h_{\mu \nu}$.
At linearized level, we have
\begin{eqnarray}
R_{\mu\nu}&=&  \partial^\rho\partial_{(\mu}h_{\nu) \rho}
-\frac{1}{2}\partial_\mu\partial_\nu h - \frac{1}{2}\square h_{\mu\nu} \;, \nonumber\\
\partial^\rho H_{\rho\mu\nu}&=&\square B_{\mu\nu}-\partial_\mu\partial^\rho B_{\rho\nu}+\partial_\nu\partial^\rho B_{\rho\mu} \;,
\label{linRH}
\end{eqnarray}
where $h = \eta^{\mu \nu} \, h_{\mu \nu}$ and
$H_{\mu\nu\rho}=3\partial_{[\mu}B_{\nu\rho]}$.
Here
we use the notation $(ab) = \frac12 (a b + ba )$
and $[ab] = \frac12 [a b - ba ]$.

The equations of motion can be consistently linearized in a region of spacetime $\Sigma$ where 
all sources of the full non-linear gravity theory are negligible in the weak field approximation. 
Thus, the linearized fields defined in this region satisfy free equations of motion with some appropriately chosen boundary conditions along $\partial\Sigma$. 
Alternatively, this boundary condition information can be encoded in terms of effective sources for the linearized fields, whose definition is extended beyond the region $\Sigma$. This technique is quite standard in the linearized gravity literature and is, in fact, convenient for our double copy construction. 
Hence, in this prescription, the linearized equations of motion for the fluctuations $h_{\mu \nu}, B_{\mu \nu}, \phi$, with effective sources $T_{\mu\nu}^{lin},j_{\mu\nu}^{(B)},j^{(\phi)}$ read: 
\begin{eqnarray}
\label{gravity_eom}
R_{\mu\nu}-\frac{1}{2}\eta_{\mu\nu}R &=&T_{\mu\nu}^{lin} \;, \nonumber\\
\partial^\rho H_{\rho\mu\nu}&=&
j_{\mu\nu}^{(B)} \;, \nonumber\\
\square \phi &=&j^{(\phi)} \;.
\end{eqnarray}

We will now go on to describe the double copy dictionary, based on two Yang-Mills one-forms $A^i$ and ${\tilde A}^{i'}$.
Here, the indices $i, i'$ denote adjoint indices.
At the linearized level, the equations of motion for $A^i$ read
\be
\partial^\mu F_{\mu\nu}^i=j_\nu^i,\quad \partial^\mu(^*F_{\mu\nu}^i)=0,\quad \partial^\mu j_\mu^i=0 \;,
\label{YMj}
\ee 
and similarly for the one-form ${\tilde A}^i$. Here, $^* F_{\mu\nu}$ denotes the dual of $F_{\mu\nu}$. In addition, there is a spectator field $\Phi_{i i'}$
satisfying
\be
\square \Phi_{ii'}=j_{ii'}^{(\Phi)}  \;.
\label{specsour}
\ee
This spectator field is needed in order to define double copy expressions,
such as 
$A_{\mu}^i \star \Phi_{ii'}^{-1} \star {\tilde A}_{\nu}^{i'}$,
that are inert under linearized non-Abelian global transformations \cite{Anastasiou:2014qba},
\begin{eqnarray}
\delta A_\mu^i &=& \partial_\mu \alpha^i + f^i_{\ jk}A_\mu^j\theta^k \;, \nonumber\\
\delta \Phi_{ii'}^{-1} &=&-f^j_{\ ik}\Phi_{ji'}^{-1}\theta^k-f^{j'}_{\ i'k'}\Phi_{ij'}^{-1}\tilde{\theta}^{k'} \;,
\end{eqnarray}
and similarly for ${\tilde A}_{\nu}^{i'}$. Here, we are considering linearized Yang-Mills theories, where 
$\alpha^i$ are local parameters and $\theta^i$ are global parameters. 
The field  $\Phi_{ii'}^{-1}$ denotes the convolution inverse of $\Phi_{ii'}$, i.e. $\Phi_{ii'}^{-1} \star
\Phi_{j j'} = \delta_{ij} \delta_{i' j'} \delta^{(4)}(x)$,
as dictated by mass dimension considerations\footnote{The fluctuations $h_{\mu \nu}, B_{\mu \nu}, \phi, A_\mu^i$ and ${\tilde A}_{\nu}^{i'}$
have mass dimension 1, as dictated by their action functionals. One can immediately see that a naive dictionary of the form $h_{\mu\nu}=A_{(\mu}^i\star\tilde{A}_{\nu)i}$ would yield the wrong mass dimension for $h_{\mu\nu}$. It is easy to check that the expression $A_{\mu}^i \star \Phi_{ii'}^{-1} \star {\tilde A}_{\nu}^{i'}$ is of mass dimension 1, as needed (note that the convolution inverse of a field $\Psi$ of mass dimension $[\Psi]=m\ $ has mass dimension $[\Psi^{-1}]=8-m $). Thus, using $\Phi_{ii'}^{-1}$, we can write a natural double copy dictionary, without the need for artificial mass-dependent coefficients. }
and the way it appears in the amplitudes double copy \cite{Cachazo:2013iea}. Its role in constructing solutions was discussed in \cite{Cardoso:2016amd}. 
In what follows,
we will omit the dependence on the spectator field in order to keep the expressions as simple as possible.
The spectator field can easily be reinstated in the double copy expressions that we will
obtain below.

Next, we write down double copy expressions for the gravitational fields  $h_{\mu \nu}, B_{\mu \nu}, \phi$ in terms of Yang-Mills configurations $A_{\mu}$ and ${\tilde A}_{\nu}$, where we suppress the dependence on the adjoint index.
Taking into account the symmetries of the fields $h_{\mu \nu}, B_{\mu \nu}$ and $ \phi$, the mass dimension of $A$ and $\tilde{A}$ (which is $1$)
and allowing for the presence of terms proportional to $1/\Box$ (c.f. (\ref{defboxi})), we express the fluctuations  $h_{\mu \nu}, B_{\mu \nu}, \phi$ as linear combinations of double copy expressions
of correct mass dimension constructed out of these ingredients,
up to (field-dependent) 
terms of the form $\partial_{(\mu} \xi_{\nu )}$ which correspond to linearized diffeomorphisms in gravity%
\footnote{
An example thereof is provided by adding to (\ref{original_dict}) a term
$$
\frac{1}{\Box} \partial_{(\mu} A_{\rho} \star \partial_{\nu)} {\tilde A}^{\rho} = 
\partial_{(\mu} \left( \frac{1}{\Box} \star \partial_{\nu)} {\tilde A}^{\rho} \right) \;. 
$$
}, and up to terms of the form $\partial_{[\mu} \Lambda_{\nu ]}$ which correspond to gauge transformations of $B_{\mu \nu}$,
\begin{eqnarray}
\label{original_dict}
 h_{\mu\nu}&=& 
 A_\mu\star\tilde{A}_\nu + A_\nu\star\tilde{A}_\mu
 -q \, \eta_{\mu\nu}\Big(A_\rho\star\tilde A^\rho-\frac1\square(\partial\cdot A)\star(\partial\cdot\tilde A)\Big) \;,\nonumber\\
  B_{\mu\nu}&=& A_\mu\star\tilde{A}_\nu - A_\nu\star\tilde{A}_\mu \;, \nonumber\\
 \phi&=& A_\rho \star \tilde{A}^\rho -\frac{1}{\square}\left(\partial \cdot A \star \partial \cdot \tilde{A} \right) \;,
\end{eqnarray}
where $\partial\cdot A = \partial_{\mu} A^{\mu}$ and $q\in \mathbb{R}$ is a priori an arbitrary constant.
Henceforth, we will take $q=1$ for calculational simplicity.
The relative coefficients between the various terms on the right 
hand side are fixed by demanding that under gauge transformations,
the combinations on the right hand side shift by amounts that can be reinterpreted as linearized local symmetry transformations
for the fields on the left hand side,
\begin{eqnarray}
\label{gravity_transforms}
\delta h_{\mu\nu}&=&\partial_\mu\xi_\nu+\partial_\nu\xi_\mu \;, \nonumber\\
\delta B_{\mu\nu}&=&\partial_\mu\Lambda_\nu-\partial_\nu\Lambda_\mu \;, \nonumber\\
\delta\phi&=& 0 \;.
\end{eqnarray}
Indeed, applying the gauge transformations $A \rightarrow A + d \alpha$ and  ${\tilde A} \rightarrow {\tilde A} + d {\tilde 
\alpha}$ on the right hand side of (\ref{original_dict}), we obtain to linear order in $\alpha$ and in ${\tilde \alpha}$ that $\phi$ is invariant and
\begin{eqnarray}
 \xi_\mu &= & \alpha\star\tilde{A}_\mu+A_\mu\star\tilde{\alpha} \:, \nonumber\\
 \Lambda_\mu&=& \alpha\star\tilde{A}_\mu-A_\mu\star\tilde{\alpha} \;.
 \label{diffgau}
\end{eqnarray}
In deriving (\ref{diffgau}), we used the convolution properties (\ref{boxgau1}) and (\ref{boxgau2}). 

We note that in the Lorentz gauge $\partial \cdot {\tilde A} =0$, the double copy expressions
(\ref{original_dict}) 
are related to those obtained in \cite{Cardoso:2016ngt,Cardoso:2016amd} in the context of the $N=2$ supergravity model $F(X) = -i X^0 X^1$, as we verify in the appendix.

The double copy dictionary (\ref{original_dict}) expresses gravitational fields in terms of Yang-Mills fields.
Similarly, we may express the
gravitational effective sources
on the right hand side of (\ref{gravity_eom}) in terms of the Yang-Mills sources in (\ref{YMj}), to obtain
\begin{eqnarray}
T^{lin}_{\mu\nu} &=& -\frac1\square\,j_{(\mu} \star \tilde j_{\nu)} \;, \nonumber\\
j_{\mu \nu}^{(B)} &=& \frac{2}{\square}j_{[\mu} \star {\tilde j}_{\nu]} \;, \nonumber\\
j^{(\phi)} &=& \frac1\square\,j_{\rho} \star {\tilde j}^{\rho} \;.
\label{sourcedict}
\end{eqnarray}
as follows. We begin by expressing the linearized Ricci tensor, given in (\ref{linRH}),
in terms of double copy data,
\begin{eqnarray}
R_{\mu\nu} &=& 
\partial^\rho \partial_{(\mu}A_{\nu)}\star\tilde A_\rho 
+\partial_\rho\partial_{(\mu}A^\rho\star\tilde A_{\nu)}
-\partial_\mu\partial_\nu A_\rho\star\tilde A^\rho \nonumber\\
&&
-\square A_{(\mu}\star\tilde A_{\nu)}
+\partial_\mu\partial_\nu \frac1\square j_\rho\star\tilde A^\rho
+\frac12\eta_{\mu\nu}j_\rho\star\tilde A^\rho \,,
\end{eqnarray}
where we used the equation of motion for $A_{\mu}$, i.e. $\Box A_{\mu} - \partial_{\mu} ( \partial \cdot A ) = j_{\mu}$.
We rewrite the second and third terms as
\begin{eqnarray}
\partial_\rho\partial_{(\mu}A^\rho\star\tilde A_{\nu)} &=&
\square A_{(\mu}\star\tilde A_{\nu)}
- j_{(\mu}\star\tilde A_{\nu)} \,,\nonumber\\[1ex]
-\partial_\mu\partial_\nu A_\rho\star\tilde A^\rho &=&
-\partial_\mu\partial_\nu \frac1\square\left((\partial\cdot A)\star(\partial\cdot\tilde A)+j_\rho\star\tilde A^\rho\right)\,,
\end{eqnarray}
to obtain
\begin{eqnarray}
R_{\mu\nu} &=&
- j_{(\mu}\star\tilde A_{\nu)} +\frac1\square \partial_{(\mu}j_{\nu)}\star(\partial\cdot\tilde A) +\frac12\eta_{\mu\nu}j_\rho\star\tilde A^\rho \,.
\end{eqnarray}
It follows that
\begin{eqnarray}
T^{lin}_{\mu\nu} = \left(R_{\mu\nu}-\frac12\eta_{\mu\nu}R \right)=
- j_{(\mu}\star\tilde A_{\nu)} +\frac1\square \partial_{(\mu}j_{\nu)}\star(\partial\cdot\tilde A)\,.
\end{eqnarray}
By also making use of the equations of motion for ${\tilde A}_{\mu}$, this may be brought into the symmetric form
\begin{eqnarray}
T^{lin}_{\mu\nu} = -\frac1\square\,j_{(\mu} \star \tilde j_{\nu)} \,.
\label{dc_graviton_source1}
\end{eqnarray}
Note that $T^{lin}_{\mu\nu}$ is conserved, i.e. $\partial^\rho T^{lin}_{\rho\mu}=0$. This is a consequence of the conservation equations $\partial^{\mu} j_\mu =0$ and $\partial^{\mu} {\tilde j}_{\mu} =0$.

Next, using (\ref{linRH}) and expressing $B_{\mu \nu}$ in terms of double copy data
(\ref{original_dict}), we obtain from (\ref{gravity_eom}),
\begin{eqnarray}
j_{\mu \nu}^{(B)} &=&
\square A_\mu \star \tilde{A}_\nu-A_\nu \star \square\tilde{A}_\mu-\partial_\mu\partial^\rho A_\rho \star \tilde{A}_\nu
+A_\nu \star \partial_\mu\partial^\rho\tilde{A}_\rho+\partial_\nu\partial^\rho A_\rho \star \tilde{A}_\mu-A_\mu \star \partial_\nu\partial^\rho\tilde{A}_\rho \;, \nonumber\\
\end{eqnarray}
which, by making use of the equation of motion for $A_{\mu}$ and ${\tilde A}_{\mu}$, can be rewritten into
\be
 j_{\mu\nu}^{(B)} =j_\mu \star \tilde{A}_\nu-A_\nu \star \tilde{j}_\mu+\partial_\nu\partial^\rho A_\rho \star 
 \tilde{A}_\mu-A_\mu \star \partial_\nu\partial^\rho\tilde{A}_\rho \;.
\ee
Then, rearranging each term in the above,
\begin{eqnarray}
j_\mu \star \tilde{A}_\nu&=& \frac{1}{\square}\left(j_\mu \star \square\tilde{A}_\nu\right) \;, \nonumber\\
-A_\nu \star \tilde{j}_\mu&=& -\frac{1}{\square}\left(\square A_\nu \star \tilde{j}_\mu\right) \;, \nonumber\\
\partial_\nu\partial^\rho A_\rho \star \tilde{A}_\mu &=& \frac{1}{\square}\left(\partial_\nu\partial^\rho A_\rho
\star \square\tilde{A}_\mu \right)  \nonumber\\
&=& \frac{1}{\square}\left(\partial_\nu\partial^\rho A_\rho \star \tilde{j}_\mu \right)
+\frac{1}{\square}\left(\partial_\nu\partial^\rho A_\rho \star \partial_\mu\partial^\tau\tilde{A}_\tau \right) \;,
\nonumber\\
-A_\mu \star \partial_\nu\partial^\rho\tilde{A}_\rho &=& -\frac{1}{\square}\left(\square A_\mu \star \partial_\nu\partial^\rho\tilde{A}_\rho\right) \;, \nonumber\\
&=& -\frac{1}{\square}\left(j_\mu \star \partial_\nu\partial^\rho\tilde{A}_\rho\right)-\frac{1}{\square}\left(\partial_\mu\partial^\tau A_\tau \star \partial_\nu\partial^\rho\tilde{A}_\rho\right) \;,
\end{eqnarray}
results in 
\be
j_{\mu\nu}^{(B)}=\frac{2}{\square}  j_{[\mu } \star \tilde{j}_{\nu]} \;.
\ee
Finally, expressing $\phi$ in terms of double copy data (\ref{original_dict}), and 
using the equation of motion for $\phi$ given in (\ref{gravity_eom}), we obtain
\begin{equation}
j^{(\phi)} = \Box A_{\rho} \star {\tilde A}^{\rho} - (\partial \cdot A) \star (\partial \cdot {\tilde A})
=
 \frac1\square\,j_{\rho} \star \tilde j^{\rho} \;,
\end{equation}
where we used the conservation equations $\partial^{\mu} j_\mu =0$ and $\partial^{\mu} {\tilde j}_{\mu} =0$.
In deriving the source dictionary ({\ref{sourcedict}), we made use 
of the convolution properties discussed earlier.

\subsection{Dictionary constraint and missing d.o.f.}\label{sec:missing dof}

Now, let us return to the source dictionary (\ref{sourcedict}), from which we infer the relation
\be
j^\phi=-T^{lin \;}_{\rho}{}^{\rho} \;.
\label{jT}
\ee
Thus, 
the sources of the graviton and dilaton are not independent!  This can be seen as a constraint on gravitational theories that
admit a double copy description at the linearized level.%
\footnote{Notice the effective sources in (\ref{gravity_eom}) are not entirely arbitrary, as they must encode information on the strongly coupled region of a classical solution of the gravity (and $B_{\mu\nu}$ and $\phi$) equations of motion. However, (\ref{jT}) cannot arise from such a requirement as it is violated by the linearization of simple solutions such as a Schwarzschild black hole.\\
Moreover, in the region where the effective sources vanish the standard on-shell counting of degrees of freedom holds, which shows that a scalar field $\phi$ is indeed part of the double copy spectrum, so that we cannot simply reabsorb $\phi$ into a redefinition of the metric perturbation.}
Note that this is a general feature of the classical double copy, and not a consequence of our set-up. To understand this, let us consider the degree of freedom counting for our map.

In the presence of arbitrary sources, 
the equations of motion do not subtract any degrees of freedom, so that the counting
of degrees of freedom (d.o.f) 
 is the same as the off-shell counting: 6 d.o.f. for $h_{\mu\nu}$, 3 for $B_{\mu\nu}$ and 1 for $\phi$. 
 Therefore, the double copy description of this theory cannot be solely based on 
 two copies of one-forms $A$ and $\tilde A$, which only contain 
 $3\times 3$ d.o.f. . The missing degree of freedom will have to be provided by the inclusion of additional
 fields.  

We note that similarly constrained gravitational systems were obtained in the amplitudes inspired construction of double copy solutions, in \cite{Luna:2016hge,Luna:2017dtq,Goldberger:2016iau,Goldberger:2017ogt,Goldberger:2017vcg}.

In the next section, we present a simple, albeit seemingly ad-hoc extension of the above double copy dictionary that includes
the missing degree of freedom.

\section{Missing degree of freedom and further sectors}

To address the missing degree of freedom,
we add sectors on the Yang-Mills side, which here we take to be sourced auxiliary fields, for simplicity,
\be
\mathcal{L}=\mathcal{L}_{YM}+\frac{1}{2}y^2-yj^{(y)} \;,
\ee
and similarly for the second copy.
In the absence of an external source, $y$ is set to zero by its equation of motion.
Also note that $y$ has mass dimension $2$.
The equations of motion for the scalar fields $y, {\tilde y}$ are then simply
\begin{eqnarray}
y &=& j^{(y)}  \;, \nonumber\\
{\tilde y} &=& {\tilde j}^{({\tilde y})} \;. 
\end{eqnarray}
Here we have again omitted adjoint indices.

Adding these sectors results in modifications of the double copy dictionary (\ref{original_dict}).
Taking into account the mass dimensions of $y, {\tilde y}$, and demanding that (\ref{gravity_transforms})
remains preserved, results in
(again up to 
terms of the form $\partial_{(\mu} \xi_{\nu )}$ which correspond to linearized diffeomorphisms in gravity,
and up to terms of the form $\partial_{[\mu} \Lambda_{\nu ]}$ which correspond to gauge transformations of 
$B_{\mu \nu}$)
\begin{eqnarray}
 h_{\mu\nu}&=& 
 A_\mu\star\tilde{A}_\nu + A_\nu\star\tilde{A}_\mu
 -\eta_{\mu\nu}\Big(A_\rho\star\tilde A^\rho-\frac1\square(
 \left(
  \partial\cdot A)\star(\partial\cdot\tilde A)  + \beta \; y \star {\tilde y} \right) \Big)
    \;,\nonumber\\
  B_{\mu\nu}&=& A_\mu\star\tilde{A}_\nu - A_\nu\star\tilde{A}_\mu \;, \nonumber\\
 \phi&=& A_\rho \star \tilde{A}^\rho -\frac{1}{\square}\left(\partial \cdot A \star 
 \partial \cdot \tilde{A} 
  + \gamma \; y \, \star {\tilde y} \right)\;,
\label{fullDict}
\end{eqnarray}
where $\beta, \gamma \in \mathbb{R}$ are arbitrary constants.  This modified dictionary produces changes
in the double copy dictionary of the sources (\ref{sourcedict}), which are straightforward to compute.
The additional term in $h_{\mu \nu}$ leads to an additional term in $R_{\mu \nu}$ given by
$-\beta \left(  \partial_{\mu} \partial_{\nu}/\Box + \frac12 \eta_{\mu \nu} \right) y \star {\tilde y}$,
and this in turn leads to a new term in $T^{lin}_{\mu \nu}$. We obtain
\begin{eqnarray}
\label{source_dict_yy}
T^{lin}_{\mu\nu} &=& -\frac1\square\,j_{(\mu} \star \tilde j_{\nu)} - \beta \left(\frac{1}{\Box} \, \partial_{\mu}
\partial_{\nu} - \eta_{\mu \nu} \right)  j^{(y)} \star {\tilde j}^{({\tilde y})}
 \;, \nonumber\\
j_{\mu \nu}^{(B)} &=& \frac{2}{\square}  j_{[\mu } \star \tilde{j}_{\nu]}  \;, \nonumber\\
j^{(\phi)} &=& \frac1\square\,j_{\rho} \star {\tilde j}^{\rho} - \gamma \,  
j^{(y)} \star {\tilde j}^{({\tilde y})} \;.
\label{sourcedicty}
\end{eqnarray}
Note that 
\begin{equation}
j^\phi + T^{lin\;}_{\rho}{}^{\rho} =  ( 3 \beta - \gamma) \, j^{(y)} \star {\tilde j}^{({\tilde y})} \;,
\label{summ}
\end{equation}
and hence, the trace of the energy-momentum tensor and the source of the scalar field $\phi$ are now disentangled.

There is, however, also a price to pay for having added 
sectors on the Yang-Mills side. Namely, we can now construct new fields on the gravitational side 
by means of the double copy, as follows,
\begin{eqnarray}
V_\mu&=A_\mu\star\tilde{y}+y\star\tilde{A}_\mu \;, \nonumber\\
W_\mu&=A_\mu\star\tilde{y}-y\star\tilde{A}_\mu \;.
\end{eqnarray}
Hence, if we do not want to enlarge the set of fields on the gravitational side of the double copy,
we will have to remove them from the set. One way of doing so consists in exploiting the non-Abelian
nature of the Yang-Mills fields on the double copy side.  To this end, we reinstate the gauge indices
of these fields.

We take $A^i_{\mu}$ and ${\tilde A}^{\tilde{j}}_{\mu}$ to be 
in the adjoint representation of a non-Abelian gauge group
$G=\tilde{G}$, while taking $y^a$ and ${\tilde y}^{\tilde{a}}$ in a different representation, for example in the fundamental representation of $G$. Then, choosing
the spectator field to be block diagonal,
\begin{equation}
\label{blockdiagspec}
\Phi^{-1} = \left(
\begin{array}{cr}
\Phi^{-1}_{i\tilde{j}} & 0\\
0 & \Phi^{-1}_{a\tilde{b}}
\end{array}
\right) \;,
\end{equation}
eliminates the fields $V_\mu$ and $W_\mu$ by virtue of $\Phi^{-1}_{i\tilde{b}}=\Phi^{-1}_{a\tilde{j}}=0$. This elimination method has already been employed in the context of the double copy in \cite{Chiodaroli:2015rdg,Chiodaroli:2015wal,Johansson:2014zca,Anastasiou:2016csv,Naculich:2014naa}.

However, when embedding our minimal solution within a larger theory we might find that we are unable to have $A_\mu$ and $y$ in different representations. Again, for simplicity, let us pick non-Abelian gauge groups
$G=\tilde{G}$,  and let's take the spectator field $\Phi^{-1}_{ij}$  to be \emph{diagonal}, so that
$V_\mu =A_\mu^ i \star  \Phi^{-1}_{ij} \star
\tilde{y}^j  + y^i \star  \Phi^{-1}_{ij} \star \tilde{A}_\mu^j $ and $W_\mu 
= A_\mu^ i \star  \Phi^{-1}_{ij} \star
\tilde{y}^j  - y^i \star  \Phi^{-1}_{ij} \star \tilde{A}_\mu^j $.
Then, we 
pick (up to gauge transformations that preserve the form of $A_\mu^i$ and of ${\tilde A}_{\mu i}$ )
\begin{eqnarray}
j_\mu^i =  (j_\mu^1,...,j_\mu^k,0,0,...) &\Longrightarrow & A_\mu^i=(A_\mu^1,...,A_\mu^k,0,0,...) \;, \nonumber\\
\tilde{j}_{\mu i} =  (\tilde{j}_{\mu 1},...,\tilde{j}_{\mu k},0,0,...)  &\Longrightarrow & \tilde{A}_{\mu i}=(\tilde{A}_{\mu 1},...,\tilde{A}_{\mu k},0,0,...) \;, \nonumber\\
j^{(y)i}=  (0,...,0,j^{(y)k+1},j^{(y)k+2},...)  &\Longrightarrow & y^i=(0,...,0,y^{k+1},y^{k+2},...) \;,\nonumber\\
\tilde{j}^{(\tilde{y})}_i =  (0,...,0,\tilde{j}^{(\tilde{y})}_{k+1},\tilde{j}^{(\tilde{y})}_{k+2},...)
&\Longrightarrow 
&\tilde{y}_i=(0,...,0,\tilde{y}_{k+1},\tilde{y}_{k+2},...) \;,
\label{removeVW}
\end{eqnarray}
in order to eliminate the fields $V_\mu$ and $W_\mu$.

At the level of counting of degrees of freedom, a more physical origin of the missing degree of freedom discussed in section~\ref{sec:missing dof} may reside in the ghost sectors of the two field theories in the double copy.
The missing degree of freedom counting problem was first addressd in a double copy-like setup by Siegel \cite{Siegel:1995px}, who noted that the total ghost number of a state described as a product of a ghost  in the left copy and another in the right copy was zero, and therefore this state lay in the physical spectrum.
  A putative physical state such as this could therefore compensate for the missing degree of freedom in a double copy description of gravity.
However,
this intriguing idea must be examined in careful detail
before it can be  offered as a putative resolution of the d.o.f
problem in a rigorous double copy description. Some work on these aspects is being addressed in \cite{Borsten:2017jpt,SD}.

\subsection{An example: the Schwarzschild black hole}

The double copy dictionary implies a correspondence between
vacua of gravity and vacua of field theories.  However, the field
equations of Einstein gravity in vacuum admit more than one static solutions:
both Minkowski space-time and the Schwarzschild solution are static solutions
of the vacuum field equations.  Conventionally, we relate the Minkowski space-time
solution to the vacuum on the field theory side. Therefore, in order to write
down a double copy dictionary for the Schwarzschild solution, we need an additional ingredient on the double copy 
side, which here is the additional sector $(y, {\tilde y})$ that we introduced above.

Let us consider the Schwarzschild solution in spherical coordinates, in the region $r > 2m$,
\begin{equation}
ds^2 = - \left(1 - \frac{2m}{r} \right) dt^2 + \left(1 - \frac{2m}{r} \right)^{-1} dr^2 + r^2 \, d\Omega^2 \;.
\end{equation}
It can be brought into isotropic form by the coordinate transformation
\begin{equation}
r = \rho \left(1 + \frac{m}{2 \rho} \right)^2 \;.
\end{equation}
The resulting line element takes the form
\begin{equation}
ds^2 = - \left( \frac{1 - \frac{m}{2 \rho}}{1 + \frac{m}{2 \rho}} \right)^2
dt^2 + \left(1 + \frac{m}{2 \rho} \right)^{4} \left(d\rho^2 + \rho^2 \, d\Omega^2 \right) \;.
\label{isotr_schwarz}
\end{equation}
In these coordinates, the horizon is at $\rho = m/2$. Introducing Cartesian coordinates $x^i$ with $\rho^2 = x^i x^i$,
and linearizing the line element (\ref{isotr_schwarz}) in $m$, we obtain
\be
ds^2 = \eta_{\mu\nu}dx^{\mu}dx^{\nu}+\frac{2m}{\rho}(dt^2+ dx^i dx^i)   \;.
\label{lin-schwarz_iso}
\ee 
Using (\ref{fullDict}), 
a double copy description of (\ref{lin-schwarz_iso}) that also ensures $B_{\mu\nu} = 0, \phi =0$ 
is obtained by picking
\begin{eqnarray}
A_{\mu}^i&=& \left(\frac{ 2m}{\rho}, 0,0,0 \right)c^i, \qquad\qquad\quad\;\; y^a=\delta^{(3)}(x) \, d^a \;,\nonumber\\
\tilde{A}_{\mu}^{\tilde{\imath}} &=&\left(\delta^{(4)}(x),0,0,0\right)\tilde{c}^{\tilde{\imath}}, \qquad\qquad\;\ \tilde{y}^{\tilde{a}}=\delta^{(4)}(x) \, {\tilde d}^{\tilde{a}} \;,\nonumber\\
\beta&=&0,\quad \gamma= 1 \;,
\label{one_centered}
\end{eqnarray}
where we have reinstated the dependence on the spectator indices, as in  (\ref{blockdiagspec}). 
The cross-terms $V_\mu,W_\mu$ vanish by virtue of the block-diagonal form of the spectator, as in (\ref{blockdiagspec}). We pick
\be 
\Phi_{AB}^{-1}=\delta^{(4)}(x)V_{AB},\qquad\qquad V_{AB}=\left(
\begin{array}{cr}
V_{i\tilde{\jmath}} & 0\\
0 & V_{a\tilde{b}}
\end{array}
\right) \;,
\ee
such that the constants of appropriate mass dimension $c^i, \tilde{c}^{\tilde{\jmath}}, d^a, {\tilde d}^{\tilde{b}}, V_{i\tilde{\jmath}}, V_{a\tilde{b}}$ satisfy
\begin{equation}
c^i \, V_{i\tilde{\jmath}} \, \tilde{c}^{\tilde{\jmath}}= 1\;,\quad 
d^a \, V_{a\tilde{b}} \, {\tilde d}^{\tilde{b}}= -2m \;.
\label{cdconst}
\end{equation}
 Finally, it is straightforward to see that the effective sources for our solution
\be
T_{\mu\nu}^{lin}=-2m\delta^{(3)}(x) \, {\rm Diag} (1,0,0,0),\qquad j_{\mu \nu}^{(B)}=j^{(\phi)}=0  
\ee
are constructed through the dictionary (\ref{source_dict_yy}) from the Yang-Mills sources
\begin{align}
j_{\mu}^i &= \big(2m\delta^{(3)}(x), 0,0,0 \big) \, c^i, 
&  j^{(y)a}&=\delta^{(3)}(x) \, d^a \;,\nonumber\\
\tilde{j}_\mu ^{\tilde{\imath}} &=
\big( \Box+\partial_0^2\,,\ \partial_0\partial_1\,,\ \partial_0\partial_2\,,\ \partial_0\partial_3\big) \delta^{(4)}(x)   
\, \tilde  c^{\tilde \imath}  \;, \hspace{-2em}
& j^{(\tilde{y})\tilde{a}}&=\delta^{(4)}(x) \,  {\tilde d}^{\tilde{a}} \;.
\end{align}

It is worth noticing that instead of setting one field theory copy to be proportional to $\delta^{(4)}(x)$, we can instead choose $\tilde{A}_{\mu}^{\tilde{\imath}} =\left(f(x),0,0,0\right)\tilde{c}^{\tilde{\imath}}$ with $f(x)$ a smooth function with well-defined convolution inverse, and also take $\Phi^{-1}_{AB} = f(x)^{-1}V_{AB}$. This guarantees that all fields are smooth configurations, without affecting the final double copy result.

Finally, let us reinstate the parameter $q$ that appears in (\ref{original_dict}) and discuss its consequences. We do so by
multiplying the terms proportional to $\eta_{\mu \nu}$ in (\ref{fullDict}) by $q$. Next, note that the
parameter $\gamma$ in the double copy expression for $\phi$ in (\ref{fullDict}) can be absorbed
into $y \star \tilde y$, which results in a rescaling of the parameter $\beta \rightarrow {\tilde \beta} =
\beta/\gamma$. Thus, in what follows, we take the double copy expression for $h_{\mu \nu}$ to depend on 
the two parameters $q$ and $\tilde \beta$. Now let us return to the linearized Schwarzschild solution 
(\ref{lin-schwarz_iso}), and consider its double copy description in terms of non-vanishing
components $A_t^i$ and ${\tilde A}_t^{\tilde i}$, as in (\ref{one_centered}). Demanding $\phi =0$
yields
\be
\frac1\square( y^a \star \Phi_{a \tilde a}^{-1} \star {\tilde y}^{\tilde a} )= - A_t^i \star \Phi_{i \tilde i}^{-1} \star {\tilde A}_t^{\tilde i} \;,
\ee
and inserting this into the double copy expression for $h_{\mu \nu}$ gives
\be
 h_{\mu\nu}=
 A_\mu^i \star \Phi_{i \tilde i}^{-1}
 \star \tilde{A}_\nu^{\tilde i} + 
 A_\nu^i \star \Phi_{i \tilde i}^{-1}
 \star \tilde{A}_\mu^{\tilde i} + q ( 1 - {\tilde \beta} ) 
 \eta_{\mu\nu} \, 
 A_t^i \star \Phi_{i \tilde i}^{-1} \star {\tilde A}_t^{\tilde i} \;.
 \ee
Demanding $h_{tt} = h_{xx}= h_{yy}= h_{zz}$, as in (\ref{lin-schwarz_iso}), 
requires
\be
q ( 1 - {\tilde \beta} ) = 1\;,
\label{qbeta}
\ee
in which case
\be
h_{tt} = h_{xx}= h_{yy}= h_{zz} = A_t^i \star \Phi_{i \tilde i}^{-1} \star {\tilde A}_t^{\tilde i} \;.
\ee
Then, choosing the fields $A_t^i, {\tilde A}_t^{\tilde i}, y^a, {\tilde y}^{\tilde a}$ as in (\ref{one_centered}),
and the constants $c^i, {\tilde c}^{\tilde i}, d^a, {\tilde d}^{\tilde a}$ as in (\ref{cdconst}), reproduces
the linearized Schwarzschild solution (\ref{lin-schwarz_iso}), for any choice of parameters
$q$ and $\tilde \beta$ satisfying (\ref{qbeta}).  
We note, however, that the value  $\tilde \beta = 0$ (and hence $q=1$) is singled out, since this is the value that allows 
a double copy description of other types of black holes, such as single-center BPS black holes in
the $F(X) = -i X^0 X^1$ model discussed in \cite{Cardoso:2016ngt,Cardoso:2016amd}.


\subsection*{Acknowledgements}

\noindent
We would like to thank Alex Anastasiou, Paolo Benincasa, Leron Borsten, Cristina C\^amara, Mike Duff,
Ricardo Monteiro, Chris White and Michele Zoccali
for helpful discussions.  
This work was partially supported by FCT/Portugal through
UID/MAT/04459/2013 and
through FCT fellowship SFRH/BPD/101955/2014 (S. Nampuri).
G. Inverso is supported by STFC consolidated grant ST/P000754/1. S. Nagy is supported by a Leverhulme Research Project Grant.

\appendix

\section{Comparing with the $N=2$ supergravity model $F(X) = - i X^0 X^1$}\label{comapringOldPaper}

In the Lorentz gauge $\partial \cdot {\tilde A} =0$, the double copy dictionary
 (\ref{original_dict}) agrees with the one \cite{Cardoso:2016ngt} derived in the context of the $N=2$ supergravity model
 based on the prepotential $F(X) = - i X^0 X^1$. This is evident for $h_{\mu \nu}$. 
We now verify that the double copy description of the fields $\phi$ and $B_{\mu \nu}$ given above is related to the one of $X^0$.
The latter is given by
\begin{eqnarray}
\label{old}
b \partial_\mu\bar{X}^0 &=& \frac{1}{2}F_{\mu\rho}^-\star\tilde{A}^\rho \nonumber\\
  &=& \frac{1}{4}F_{\mu\rho}\star\tilde{A}^\rho+\frac{i}{8}\varepsilon_{\mu\rho\alpha\beta}F^{\alpha\beta}\star\tilde{A}^\rho \nonumber\\
&=& \frac{1}{4}\partial_\mu\left(A_\rho\star\tilde{A}^\rho\right)+\frac{i}{4}\varepsilon_{\mu\rho\alpha\beta}\partial^\alpha A^\beta\star\tilde{A}^\rho \;.
\end{eqnarray}
In the Lorentz gauge, we have $\phi = A_{\rho} \star {\tilde A}^{\rho}$, and hence
it follows immediately that
\be
\phi
=4b \, {\rm Re} \left(\bar{X}^0 \right) \;,
\ee
up to a constant. Using
\begin{eqnarray}
  b \, \varepsilon^{\mu abc} \, \partial_\mu Im\left(\bar{X}^0\right) =-3!\partial^{[a}A^b\star\tilde{A}^{c]} \;,
\end{eqnarray}
as well as $B_{\mu\nu}=A_\mu\star\tilde{A}_\nu - A_\nu\star\tilde{A}_\mu$, 
we obtain the relation
\be
H_{\mu\nu\rho}=3\partial_{[\mu}B_{\nu\rho]}=4b \, \varepsilon_{\mu\nu\rho\sigma}\partial^\sigma Im\left(\bar{X}^0\right) \;.
\ee

\providecommand{\href}[2]{#2}\begingroup\raggedright\endgroup

\end{document}